# Flow and equation of state in heavy-ion collisions

P. Danielewicz[a] *

[a] National Superconducting Cyclotron Laboratory and
Department of Physics and Astronomy, Michigan State University,
East Lansing, Michigan 48824, USA

The status of flow in heavy-ion collisions and of inference of hadronic-matter properties is reviewed.

## 1. INTRODUCTION

Collective flow is a motion characterized by space-momentum correlations of dynamic origin. It is of interest in collisions because it may tell us about pressures generating that motion and about the equation of state (EOS) and other properties of the strongly-interacting matter. The flows that have been identified thus far are radial, sideward, and elliptic. The conclusions on the properties are normally based on comparisons of data to transport-model simulations.

## 2. RADIAL EXPANSION

The collective radial expansion is often assessed by looking for deviations of momentum distributions, especially transverse, from thermal. The momentum distributions are commonly described in terms of the simple Siemens-Rasmussen [1] formula, or its derivatives,

$$\epsilon \frac{dN}{d\mathbf{p}} \propto \mathrm{e}^{-\gamma\epsilon/T} \left\{ \frac{1}{pv} \left(\gamma\epsilon + T\right) \sinh \frac{\gamma pv}{T} - \cosh \frac{\gamma pv}{T} \right\} . \tag{1}$$

A safer assessment of the radial flow is by comparing spectra, or average energies, of particles with different mass. The higher the mass, the stronger is the effect of collective expansion and flatter the distribution (more spread-out by the collective velocity), and higher the average energy. Transverse distributions flattening with the particle mass have been seen in central collisions of heavy nuclei at beam energies ranging from below 100 MeV/nucleon to above 100 GeV/nucleon, cf. Fig. 1. The larger-mass distributions are sharper in the lighter than in the heavier system at the high energy, indicating a weaker collective expansion in the lighter system.

With regard to the dependence of the strength of collective expansion on direction relative to the beam axis, the FOPI Collaboration at GSI investigated cross sections for the energy ratio $ERAT = E_\perp/E_\parallel$. A value of $ERAT = 2$ is consistent with isotropy given that there are two transverse degrees of freedom and one longitudinal. A value $ERAT <$

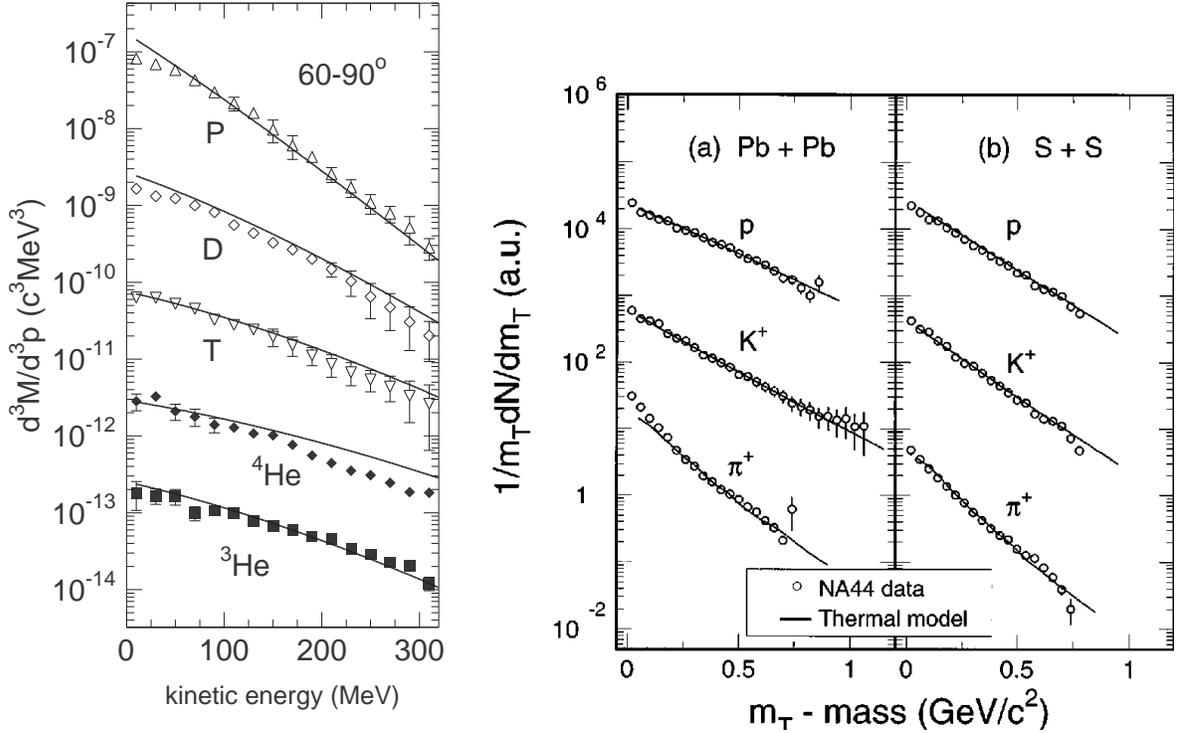

Figure 1. Transverse momentum distributions in central collisions at 250 MeV/nucleon (left) and at 158 GeV/nucleon (right) from the measurements of the FOPI [2] and of the NA44 [3] Collaborations, respectively.

2 indicates transparency and $ERAT > 2$ – a splash perpendicular to the beam axis. Taking into account efficiency cuts, from their studies of $ERAT$, the FOPI Collaboration concluded that the $b = 0$ Au + Au collisions with strong radial expansion are consistent with isotropy at 250 MeV/nucleon [4]. This may be contrasted with results of E802 Collaboration obtained in central Au + Au collisions at 10.7 GeV/nucleon [5]. When the measured [5] transverse-momentum distribution of protons at midrapidity, characterized by an effective temperature of 215 MeV, is rotated in the direction of the beam axis and then transformed to a rapidity distribution, it is found to be twice as narrow as the measured proton rapidity distribution. This indicates that the nuclei at 10.7 GeV/nucleon do not come to a complete stop even in Au + Au. As the spectra at one rapidity and particle ratios are well described within the local equilibrium model, it is apparent, though, that a local equilibration takes place during the stopping process and is nearly complete by the system freeze-out.

For a system of particles that freeze out at a common temperature and at a common velocity field, the mean particle energy should increase linearly with particle mass, for larger particle masses. Deviations from a linear behavior can offer a direct glimpse at a temporal and, possibly, spatial development of the collective expansion. Thus, e.g. the relatively low average transverse energy of $S = 3$ $\Omega$ baryons in central 158 GeV/nucleon

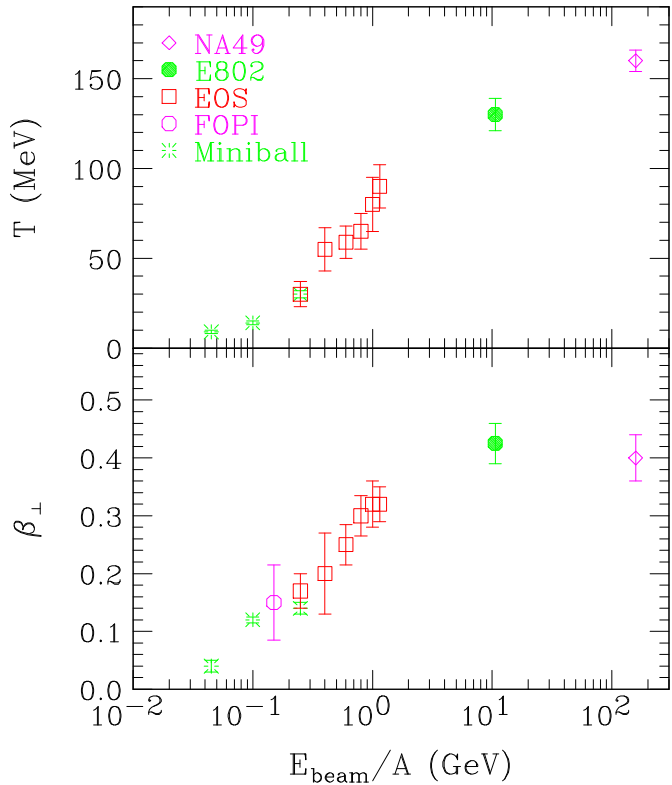

Figure 2. Excitation functions of transverse temperature (upper panel) and transverse collective velocity (lower panel) at midrapidity in heavy systems.

Pb + Pb collisions appears to indicate [6], simultaneously, a relatively early freeze out of these baryons and low values of collective energy early on during system development. Low values of the average c.m. energy of heavier intermediate mass fragments in central Xe + Sn collisions at 50 MeV/nucleon [7] appear to indicate, on the other hand, a late-time emission of these fragments and low values of collective energy towards the end of the system development when pressure has carried out most work and has decreased.

Figure 2 displays excitation function of transverse temperature and of velocity in heaviest systems. The velocity saturates at AGS energies, possibly due to meson production and progressing transparency.

Of interest is the possible use of the radial expansion in the determination of EOS. It must be remembered that the separation into the collective and thermal energies occurs at freeze-out, when collisions become infrequent. Let us consider first the situation at low energies. If the EOS is soft and pressure low, the expansion is slower than for a stiff EOS, but then one just needs to wait longer for same observable values to emerge at freeze-out. To tell the difference, one needs some timing device. As such devices might serve the persistence of longitudinal motion at high energies or the early strange particle emission. I am unaware of any use of these, though, so far in the EOS determination.

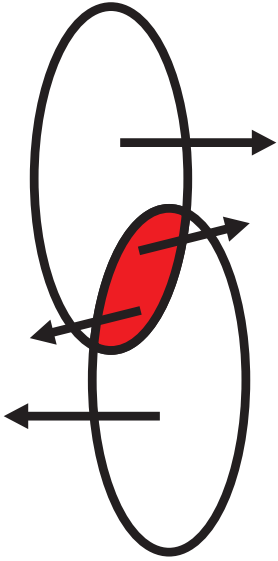

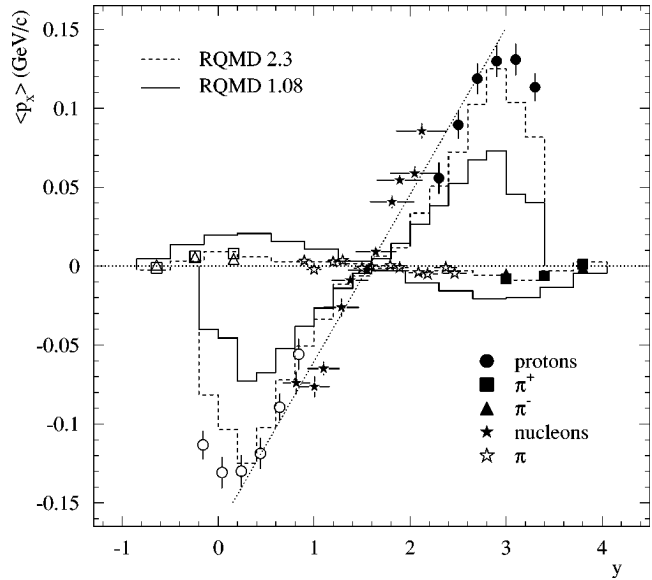

Figure 3. In-plane particle deflection.

Figure 4. Average in-plane transverse momentum component as a function of rapidity in central Au + Au collisions at 10 GeV/nucleon [8].

## 3. SIDEWARD FLOW

Sideward flow is a deflection of forwards and backwards moving particles, away from the beam axis, within the reaction plane. The situation in reactions is schematically illustrated in Fig. 3. For the compressed and excited matter in a central region it is easier to get out to the vacuum on one side of the beam axis than on the other. Eagerness to get out will be enhanced by high generated pressure but also by the momentum dependence of mean fields before the equilibration takes place. The ability to get out depends on the inter-particle cross sections.

The sideward flow is often represented in terms of the mean in-plane component of transverse momentum at a given rapidity, $\langle p^x(y)\rangle$, and additionally quantified in terms of the derivative at the midrapidity, see Fig. 4:

$$F_y = \frac{d\langle p^x\rangle}{dy} \qquad \text{or} \qquad F = \frac{d\langle p^x\rangle}{d(y/y_B)}. \tag{2}$$

The normalization of the rapidity to the beam in the derivative enhances, somewhat artificially, the strength of dynamic effects at high energies relative to low.

In transport models, it is directly observed that the production of sideward flow is shifted towards the high density phase [9] as compared to the radial flow [10]. The sideward flow thus has more potential in the EOS determination than the radial flow. The flow excitation function is represented in Fig. 5 and the flow is seen to be maximal between 0.1 and 10 GeV/nucleon.

Let me express at this point my prejudices with regard to the quark-gluon phase searches. In the region of the transition, a drop in the speed of sound or in pressure relative to energy density is expected. At the temperatures such as found for the transition in baryonless matter in lattice calculations, the hadron density is of the order of twice or so as in normal nuclear matter. Given that the transition is associated with hadrons pushing out the nonperturbative vacuum from their region, the transition is to be expected at comparable densities no matter whether the hadron density is increased through raising temperature or through compression. Densities required for the transition are reached at the AGS energies and the traces of the transition should be looked for, in that energy domain, in the flow generated by the pressure in the compressed matter.

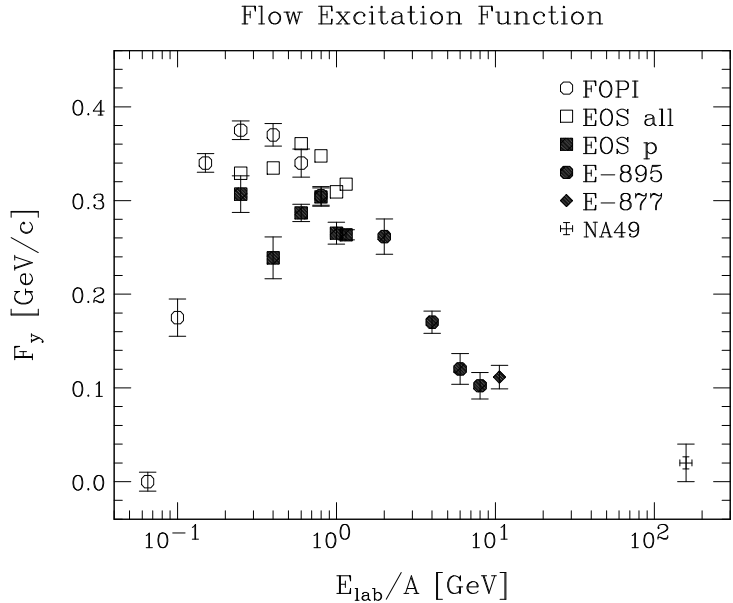

Figure 5. Excitation function of sideward flow in central collisions of heavy nuclei.

## 4. SECOND-ORDER OR ELLIPTIC FLOW

The elliptic flow is typically studied at midrapidity and quantified in terms of $v_2$:

$$v_2 = \langle \cos 2\phi \rangle \qquad (\, v_n = \langle \cos n\phi \rangle \,) \,, \qquad (3)$$

where $\phi$ is the azimuthal angle relative to the reaction plane. The second-order flow may offer a better chance for the EOS determination than the first-order sideward flow, because it involves less of the uncertainties in the opposing streams of matter moving past each other. Typical azimuthal patterns at midrapidity may be seen in Fig. 6.

At AGS energies the elliptic flow results from a competition between the early squeeze-out when compressed matter tries to move out in the direction perpendicular to the reaction plane and the late-stage in-plane emission associated with the shape of the participant zone [12], cf. Fig. 7. The squeeze-out contribution to the elliptic flow depends, generally, on the pressure $p$ built-up early on, compared to the energy density $e$ playing the role of a mass, and on the passage time for the spectators (for a more thorough discussion see [13]). When the heated matter is exposed to the vacuum in the transverse direction, a rarefaction wave moves in putting the matter into motion. The time for developing the expansion is then $R/c_s$, where $R$ is the nuclear radius and $c_s = \sqrt{\partial p/\partial e}$ is the speed of sound. The passage time for spectators, on the other hand, is of the order of $2R/(\gamma_0 v_0)$, where $v_0$ is the spectator c.m. velocity. The squeeze-out contribution to the elliptic flow should then reflect the time ratio

$$\frac{c_s}{\gamma_0 v_0} . \qquad (4)$$

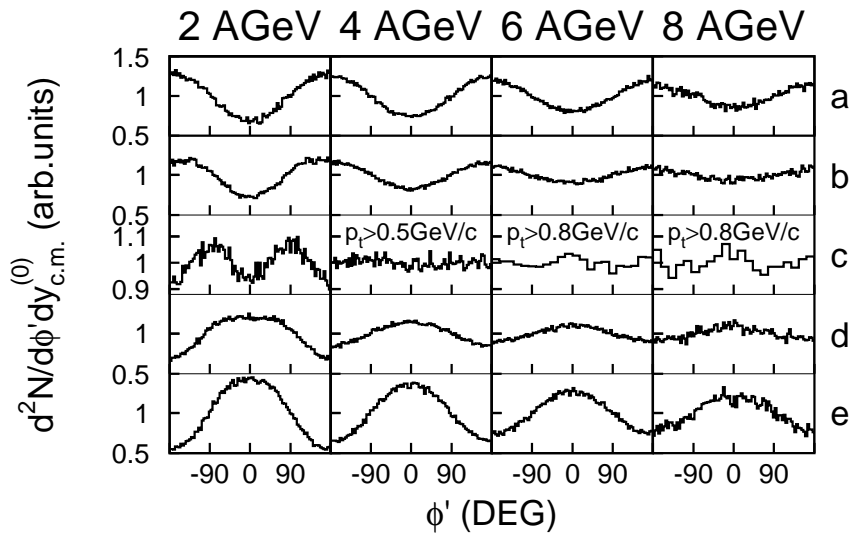

Figure 6. Azimuthal distributions, with respect to the reconstructed reaction plane, of protons emitted from semicentral Au + Au collisions [11] in the rapidity intervals of $-0.7 < y/y_{Beam} < -0.5$ (a), $-0.5 < y/y_{Beam} < -0.3$ (b), $-0.1 < y/y_{Beam} < 0.1$ (c), $0.3 < y/y_{Beam} < 0.5$ (d), and $0.5 < y/y_{Beam} < 0.7$ (e).

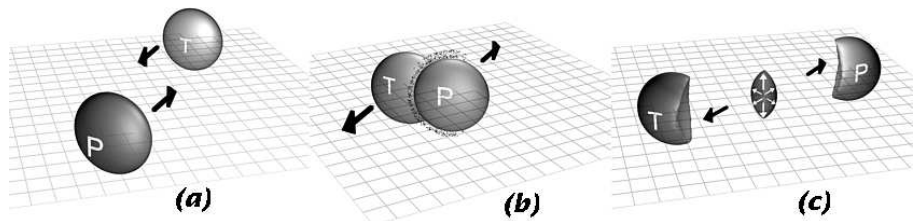

Figure 7. Collision of two Au nuclei at relativistic energies. Time shots are shown for an instant before the collision (a), early in the collision (b), and late in the collision (c).

The result (4) gives hope that any changes in the speed of sound associated with a phase transition might be revealed in the variation of the elliptic flow. Overall, the squeeze-out contribution should decrease as a function of energy with the flow becoming positive, $v_2 > 0$. The elliptic-flow excitation function is displayed in Fig. 8. It is seen that, indeed, while dominated by squeeze-out at moderate energies, the elliptic flow becomes positive at high energies. Whether or not any changes in $v_2$ might be associated with any phase transition requires comparisons to transport-model calculations to assess the magnitude of such possible changes.

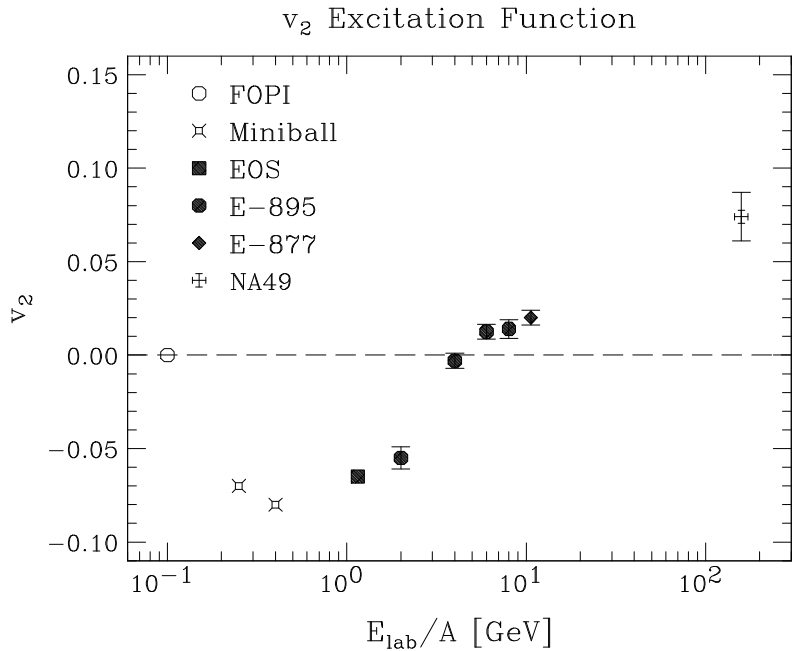

Figure 8. Elliptic-flow excitation function.

So far I discussed directional flows for nucleons and baryon clusters. These are the most abundant particles strongly sensitive to a collective motion within a colliding system. Flow of other particles, such as of pions [14] or kaons, may provide complementary information to that emerging from baryon flow, on geometry and on mean fields acting on the different particles.

## 5. TRANSPORT-MODEL COMPARISONS

The Boltzmann equation for the phase-space distribution $f$,

$$\frac{\partial f}{\partial t} + \frac{\partial \epsilon_\mathbf{p}}{\partial \mathbf{p}} \frac{\partial f}{\partial \mathbf{r}} - \frac{\partial \epsilon_\mathbf{p}}{\partial \mathbf{r}} \frac{\partial f}{\partial \mathbf{p}} = I \,, \tag{5}$$

has been the workhorse of transport reaction-simulations from moderate to high energies. The equation accounts for changes in the distribution due to the motion of particles in space (with $\partial \epsilon / \partial \mathbf{p}$ being velocity), due to acceleration under the average force from other particles (with $-\partial \epsilon / \partial \mathbf{r}$ being the force), and due to collisions (with $I$ being collision rate). The single-particle energies $\epsilon$ are related to EOS and, unfortunately, EOS is not the sole unknown quantity for the strongly-interacting medium ..., even the absence of quark-gluon plasma.

Difficulties in addressing reactions are faced because in-medium cross sections are not known. Nucleon mean field generally depends on momentum. Fields with different momentum-dependence may yield similar EOS but produce different forces out of equilibrium. The momentum dependence can compete with a stiffness of EOS in generating the directional flow. It may seem obvious that, if one is after the EOS, i.e. dependence of energy or pressure (momentum flux) on density and temperature, then a model used in the comparisons to data should have a well-defined and conserved energy-momentum flux tensor. Practice is different, at times due to technical reasons. Going beyond hadronic degrees of freedom, questions arise of how to represent the quark-gluon EOS in a simulation. Should one treat the quark-gluon phase within the hydrodynamic model and combine that model with the transport model for hadrons? Should one follow hadron dynamics emulating at high temperatures and densities the quark-gluon plasma?

The difficulties may be tackled in steps. Thus, at low energies the in-medium cross-section modifications may be assessed by using the so-called longitudinal momentum transfer and by using $ERAT$. At high energies, the distributions $dN/dy$ and $dN/d\,m_\perp$ could be used, but these also depend on EOS.

The momentum dependence of the nucleon mean field is known from nucleon-nucleus scattering experiments, but, up to now, has not been directly demonstrated in nucleus-nucleus collisions. A strong momentum dependence may generate a strong flow such as due to a stiff EOS. Effective competition between the two sources of flow may, however, take place only at low impact parameters when reached hadron densities are large. As reactions become more peripheral, the densities drop while momentum dependence stays on. Thus, effects of momentum dependence should become cleanly separated at high impact parameters. In addition, of course, one may expected enhanced effects at high momenta relative to the matter. Useful with that respect data have been taken by the KaoS Collaboration [15] and a comparison of transport-model calculations to one set of the data is shown in Fig. 9.

The three lines for $m^*/m = 1$ represent results from a transport-model calculation with a stiff and soft EOS and no momentum dependence, and without mean field, respectively. Other represented calculations correspond to different parametrizations of the momentum dependence. The two lines for $m^*/m = 0.70$ correspond to stiff and soft EOS, respectively, with an identical momentum dependence. It is apparent, that the elliptic flow in semiperipheral collisions and at high momenta gives evidence for the momentum dependence of the mean field. Moreover, the flow allows to establish details in the momentum dependence. Strong evidence for the momentum dependence is present in the KaoS data [15] up to 1 GeV/nucleon. One issue with the momentum dependence determined from the flow, in the semiperipheral collisions, is that the dependence cannot be quite utilized in central collisions at the same energy but rather at lower, because of the densities reached.

While I began to present results, I did not mention any details of the employed transport, Boltzmann-equation, model (BEM). The model is formulated within the relativistic Landau theory [16]. The degrees of freedom are nucleons, pions, delta and $N^*$ resonances. The single-particle energies and EOS are specified by parametrizing volume energy density in terms of phase-space distribution functions. Two types of parametrizations are used, in terms of scalar and vector density for momentum-independent and momentum-dependent fields, respectively. In the scalar parametrization, the baryon single-particle energies are

$$\epsilon_X(p, \rho_s) = \sqrt{p^2 + (m_X + U(\rho_s))^2}\,,$$ (6)

where $\rho_s = \sum_X \int d\mathbf{p}\, \frac{m_X(\rho_s)}{\sqrt{p^2 + m_X^2(\rho_s)}}\, f_X(\mathbf{p})$ and

$$U(\xi) = \frac{-a\,\xi + b\,\xi^\nu}{1 + (\xi/2.5)^{\nu - 1}}\,,$$ (7)

with $\xi = \rho_s/\rho_0$. In the vector model, in the frame where baryon flux vanishes, $\mathbf{J} = \sum_X \int d\mathbf{p}\, f_X(\mathbf{p})\, \frac{\partial \epsilon_X}{\partial \mathbf{p}} = 0$, the energies are

$$\epsilon_X(p, \rho) = m_X + \int_0^p dp'\, v_X^* + \rho \left\langle \int_0^p dp'\, \frac{\partial v^*}{\partial \rho} \right\rangle + U(\rho)$$ (8)

and

$$v_X^*(p, \xi) = \frac{p}{\sqrt{p^2 + m_X^2 \left/ \left(1 + c\, \frac{m_N}{m_X}\, \frac{\xi}{(1 + \lambda\, p^2/m_X^2)^2}\right)^2\right.}}\,.$$ (9)

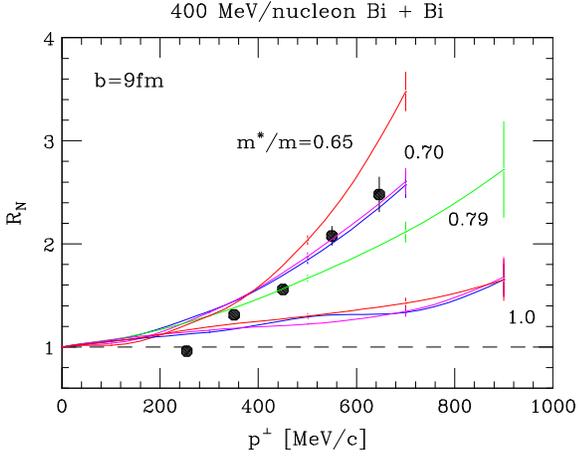

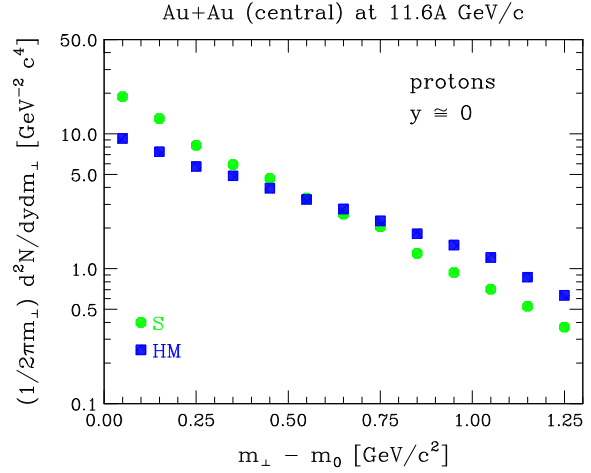

Figure 9. Out-of-plane to in-plane ratio $R_N = (1 - 2\,v_2)/(1 + 2\,v_2)$ for protons emitted at midrapidity from 400 MeV/nucleon Bi + Bi reactions, as a function of transverse momentum. Symbols represent data of Ref. [15] while lines represent transport-model calculations.

Figure 10. Proton transverse-mass distribution in central Au + Au collisions at 10.7 GeV/nucleon for soft momentum-independent (circles) and hard momentum-dependent EOS (squares).

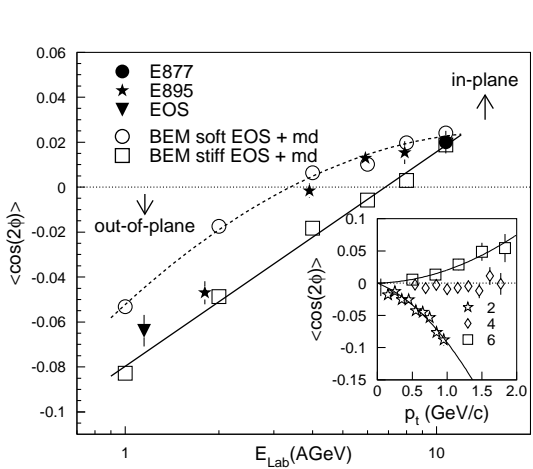
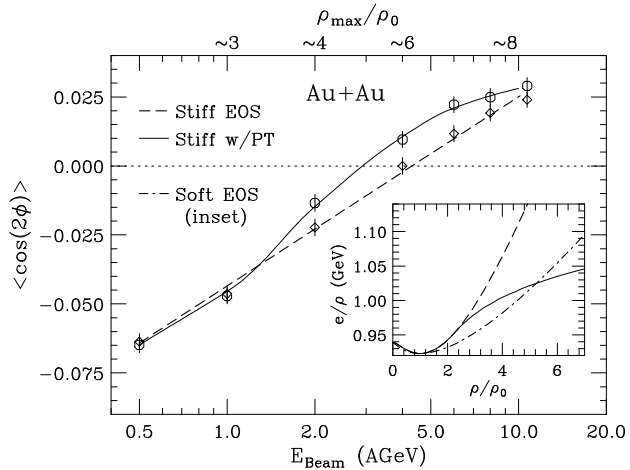

Figure 11. Elliptic-flow excitation function for Au + Au. The data are from Refs. [11], [18], and [19]. Calculations have been carried out for the soft and the stiff EOS with momentum dependence.

Figure 12. Calculated elliptic flow excitation functions for Au + Au. The diamonds represent results obtained with a stiff EOS. The circles represent results obtained with a stiff EOS and with a second-order phase transition. The lines guide the eye. The inset shows energy per baryon vs. baryon density for different EOS.

The effective mass at normal density for the momentum-dependent fields is taken equal to $m_N^* = 0.65\,m_N$. Results will be shown for incompressibility values of $K = 210$ and 380 MeV, and for an EOS with a second-order phase transition.

The rapidity and transverse-mass distributions from the model exhibit some sensitivity to EOS, cf. Fig. 10 (see also Ref. [17]). While the best agreement with data at 10.7 GeV/nucleon [5] is obtained for soft EOS or without mean field, the distributions are also sensitive to the details of energetic elementary cross sections.

With regard to the utility of the elliptic flow in the determination of EOS, Fig. 11 shows BEM results for soft and stiff EOS with momentum dependence. The softer EOS, with a lower speed of sound, produces more positive $v_2$ as expected on the basis of (4). Figure 12 shows next the expected change in the elliptic-flow excitation function in the presence of a weak phase transition. As the EOS softens with increase of energy and of density, $v_2$ moves up. It should be noted that the data in Fig. 11 exhibit an upward movement of $v_2$ between 2 and 4 GeV/nucleon beam energy, such as expected for a phase transition, at baryon densities above $4\,\rho_0$. In deciding whether the movement could indeed signal the phase transition, one would need to correlate the information on other observables sensitive to EOS, such as sideward flow and transverse-momentum distributions. Given that those observables are quite sensitive also to cross sections for different processes, the latter must be treated with much care within a model.

## 6. SUMMARY AND OUTLOOK

Flow observables are important tools for investigating properties of hadronic matter in energetic collisions. Data from semiperipheral collisions, at 1 GeV/nucleon and below, give evidence on the magnitude of the momentum dependence of mean fields in excited matter. Data from semicentral collisions at AGS energies point to a softening in the hadronic EOS above 2 GeV/nucleon or above four times the normal nuclear density. New data on sideward flow at AGS energies, with smaller errors, as well as excitation functions for more mundane single-particle observables could help to clarify whether one might, indeed, deal with a phase transition. Using variation of density with impact parameter, a possibility might be to study the variation of flow with impact parameter at one energy above the transition in central collisions, i.e. any energy $\gtrsim$ 4 GeV/nucleon.

## ACKNOWLEDGEMENT

This work was partially supported by the National Science Foundation under Grant PHY-9605207.